# Autonomous Optimization of an Organic Solar Cell in a 4-dimensional Parameter Space

Tobias Osterrieder, Frederik Schmitt, Larry Luer, Jerrit Wagner, Thomas Heumüller, Jens Hauch and Christoph Brabec

## Abstract

Optimizing solution-processed organic solar cells is a complex and challenging task due to the vast parameter space in organic photovoltaics (OPV). Classical Edisonian or one-variable-at-a-time (OVAT) optimization approaches are laborious, time-consuming, and may not find the optimal parameter set in multidimensional design spaces. To tackle this problem, we demonstrate here for the first time artificial intelligence (AI) guided closed-loop autonomous optimization for fully functional organic solar cells. We empower our LineOne, an automated materials and device acceleration platform with a Bayesian Optimizer (BO) to enable autonomous operation for solving complex optimization problems without human interference. The system is able to fabricate and characterize complete OPV devices and navigate efficiently through the design space spanned by composition and processing parameters. In addition, a Gaussian Progress Regression (GPR) based early prediction model is employed to predict the efficiency of the cells from cheap proxy measurements, in our case, thin film absorption spectra, which are analyzed using a spectral model based on physical properties to generate microstructure features as input for the GPR. We demonstrate our generic and complete autonomous approach by optimizing composition and processing conditions of a ternary OPV system (PM6:Y12:PC70BM) in a four-dimensional parameter space. We identify the best parameter set for our system and obtain a precise objective function over the whole parameter space with a minimal number of samples. We demonstrate autonomous optimization of a complex opto-electronic device within 40 samples only, whereas an Edisonian approach would have required about 1000 samples. Even larger acceleration factors are expected for higher dimensional parameter spaces. This raises an important discussion on the necessity of autonomous platforms to accelerate Material science.

# Introduction

Organic Photovoltaic (OPV) technology is highly attractive for building integrated PV [1], consumer ware [2], and greenhouse applications [3] due to its unique properties, including low weight, flexibility, and semi-transparency [4]–[7]. Additionally, OPV efficiencies have increased significantly over the last few years, approaching the 20 % efficiency milestone [8]. However, optimizing OPV devices is usually a time and workforce-intensive process due to the vast parameter space, including material, solution, and process parameters [9], leading to a large number of experiments required to find the optimal configuration. Furthermore, those parameters are often correlated, which makes them unfeasible for tedious and expensive trial-and-error methods [8], [10]. Historically, it takes about 15-20 years for a new technology to be transferred from discovery in a lab to industrial applications [11].

LineOne was first built as a Device Acceleration Platform (DAP) with the intention to accelerate this transfer significantly by utilizing automated high-throughput experiments platforms and AI-guided sample selection [12]. Here, for the first time, we demonstrate an autonomous optimization of ternary OPV devices, using a simple device structure and green solvents to enable the transfer to industrial pilot lines with minimal adjustments.

High-Throughput Materials Acceleration Platforms (MAPs) manage to produce a large number of samples in short time frames and are already successfully deployed for biomedical research [13], drug discovery [14], biology [15], and organic chemistry [16]. In material science, they are used for growing carbon Nanotubes [12], synthesizing colloidal crystals [13], and discovering new materials for organic light-emitting diodes (OLEDs) [14], catalysts [15], [16], batteries [17], [18], and photovoltaics [19], [20]. The community has acknowledged the ability of MAPs to create large, reproducible data sets, thus potentially accelerating material science significantly for several years [11]. However, the power of MAPs remains limited if they are not connected to corresponding high throughput characterization and application platforms like a DAP. Therefore, it is necessary to fuse MAPs with DAPs into an AMADAP, combining both advantages in one Automated Material and Device Acceleration Platform. Combining these concepts requires well defined interfaces and, most importantly, autonomous operation which can cover material and devices optimization. In this manuscript we describe how to evolve an automated device processing line into an autonomous operating device optimization line.

AMADAPs are set to generate large and unique reference data libraries. Similar to libraries like the Materials Genome Initiative [17], the High Throughput Experimental Database [18], and the Materials Experiments and Analysis Database [19], which collect and store large amounts of experimental and computational data to accelerate the discovery of new materials. For efficiently analyzing the obtained data, machine learning (ML) tools have become more popular among the scientific community, powered by the increased availability of open-source and user-friendly toolboxes [20]–[24]. ML provides a mapping between input and target properties in multidimensional parameter spaces [25]. Moreover, Quantitative Structure-Property Relationship (QSPR) models are based on the underlying physics and assist researchers in deepening their understanding of the investigated systems [26], [27]. In the field of OPV, Lee created a model based on a data set of 124 fullerene derivatives based ternary OSCs, which provided insights into the vital role of the Donor LUMO [28]. Wu et al. trained a Random Forest (RF) model on 565 donor-acceptor pairs extracted from the literature and determined a novel pair that achieved an impressive efficiency of 16.5 % [29], [30]. Recently our group successfully predicted the efficiency of PM6:Y6 OPVs using features obtained from absorption measurements with a Gaussian Process Regression (GPR) [31].

Utilizing AMADAPs together with sequential active learning algorithms creates self-driving labs, able to autonomously navigate and identify the optimal parameter set in multidimensional design spaces. Those closed-loop approaches are anticipated to increase the throughput and precision of experimental data generation [32] while freeing the researcher from laborious, repetitive tasks [33] and helping them to identify relevant parameters reliably, reducing the number of "unnecessary" experiments. Most commonly, a Bayesian Optimizer is incorporated to sample the design space efficiently and is currently the leading method for the exploration of low dimensional parameter spaces [34]–[39]. The algorithm chooses the following sample points based on an acquisition function that, by design, trades off exploration and exploitation of the search space. Therefore, the recorded data also includes "failed" experiments, which is essential for training ML models and is in contrast with the current publication bias towards good-performing samples [33], [40], [41]. Self-driving Labs have been successfully demonstrated in multiple applications, e.g., searching for improved photocatalysts for hydrogen production from water [42], new synthesis conditions for metallic [43], flow synthesis of organic compounds [44], and battery research [45], [46]. In organic photovoltaics, Bayesian optimization was used by our group to optimize environmental photo-stable multicomponent polymer composites for photovoltaic conversion [47] and by MacLeod et al. to successfully optimize the hole mobility of hole transfer layers [48]. However, while individual process conditions have been optimized, the full optimization of complete functional devices has never been demonstrated.

Herein we demonstrate for the first time the autonomous optimization of an organic solar cell by a self-driving lab. We show that our system does identify the optimal parameter set and quantifies the importance of the respective parameters in one precise objective function. Furthermore, we couple a GPR model, predicting device efficiency at the hand of an absorption spectrum, to the BO [31]. Operating AMADAP in that mode allows to optimize the performance of solar cells purely from absorption measurements and has the potential to reduce the optimization cycle from a few hours to a few minutes. We propose that integrating two AI approaches with fast proxy measurements is a unique strategy for further accelerating device optimization.

## Results and Discussion

We conducted two independent optimization experiments to demonstrate our system, using the data from both experiments for the early efficiency GPR prediction. In the first experiment, we limited the optimization of our ternary active layer (PM6:Y12:PC70BM) to a two-dimensional parameter space (2D), and in the second experiment, we optimized the ratios, concentration, and spin speed simultaneously in a four-dimensional parameter space (4D). The goal of both experiments was to maximize the power conversion efficiency (PCE) of the system. The workflow is shown in Figure 1. Orchestration of the devices and robots in LineOne is accomplished by an in-house developed framework [12], which can be controlled via a user-friendly web interface. BO, GPR, LineOne, and Database communicate with each other through customized APIs. All data and metadata, including basic data on the raw materials, process conditions, environmental conditions such as atmospheric values, and of course, the characterization data of the measurement devices, are stored in the database together with all relevant and accessible metadata. This enables to start and restart the optimization procedure either with no pre-existing input or with any available data sets from previous campaigns as starting point.

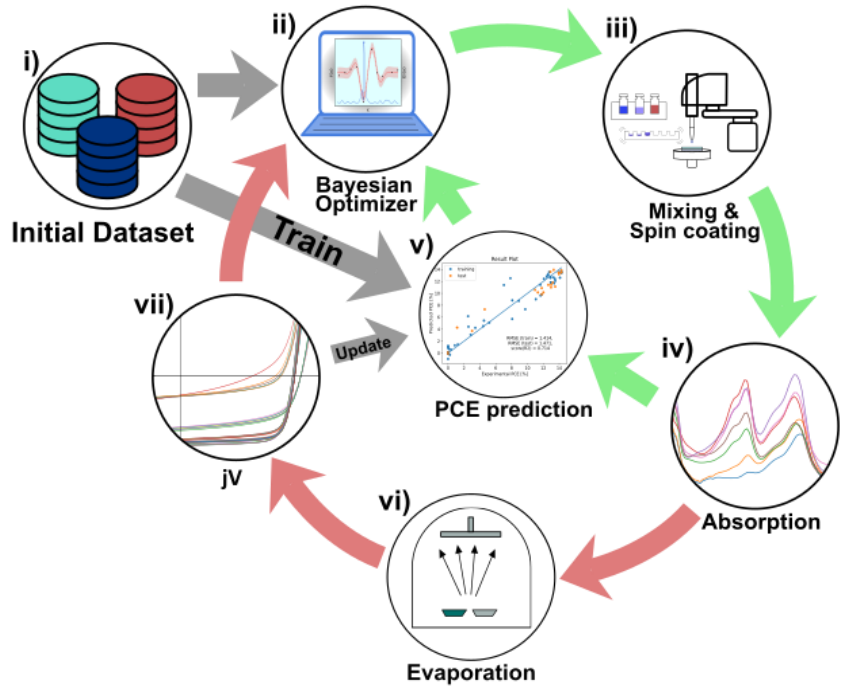

**Figure 1: Closed-Loop Optimization using UV-Vis data to predict PCE**. i) The Experiment starts with creating an initial data set by producing and characterizing a small set of devices with different parameters obtained by LHS. The Data is used to train the Bayesian Optimizer (BO) and the early PCE prediction model. ii) The BO then suggests the parameters of a batch of new samples. iii) The solutions of the active layers are automatically mixed and subsequently spin-coated onto the substrates. iv) The data from UV-Vis measurements are used to predict the efficiency using our early prediction tool. v) The predicted efficiencies are fed to the BO, which suggests new samples based on all previous data. vi) Full devices are produced by evaporating HTL and electrodes on top, and vii) the efficiencies of the complete devices are measured. The measured efficiencies are used to retrain both models to increase their accuracy.

To initialize the experiment, we created an initial data set, utilizing Latin Hypercube Sampling (LHS), containing 28 different samples and measuring absorption and jV-Characteristics of the cells. For a typical optimization procedure, we leverage the ability to produce multiple cells proposed by LHS in parallel with LineOne, which has the capacity to produce up to 80 substrates with a total of 480 cells daily. Most importantly, this data set is used to train a GPR model to predict the efficiency of the samples using features obtained by deconvoluting the absorption spectra. In an earlier publication, we showed that this approach is able to predict efficiencies of PM6:Y6 reliably and moreover provides deep insight into the structure-property relationship between the morphology and electronic features [31]. Next, LineOne feeds the obtained data to the Bayesian optimizer to initiate the closed-loop optimization process. Using LHS to calculate the parameter sets for the initial samples ensures good coverage of the design space, decreasing the number of iterations needed for optimization. For the Bayesian Optimizer, we used the open-source scikit-optimize Python package, with a Matern 5/2 Gaussian Process Kernel as the surrogate function and the "gp_hedge" acquisition function, which probabilistically chooses one of the three acquisition functions, lower confidence bound (LCB), expected improvement (EI), and probability of improvement (PI), selecting the one that yields the best gain in each iteration [49]. The BO proposes a batch of seven new samples in each iteration, using a constant liar strategy [50], while the eighth substrate on the carrier is used as a reference device. The selection of the next batch is a trade-off between exploitation and exploration of the design space, which leads to a fast and sample-efficient optimization. The parameter sets of the new samples are automatically sent to LineOne to initiate the follow-up run, which starts by mixing, spin-coating, and annealing the active layer. The absorption spectra of each layer are measured at six different locations of the substrate and spectrally modeled to extract the energetic and microstructural features that are used to predict the efficiency of the cells with GPR [31] pre-trained with the initial data set from the LHS run. The predicted values are fed back to the BO. This early prediction process reduces the time for one iteration from 200 to approximately 90 minutes since film processing is about three times faster than device processing. The major time-limiting step in the film processing is mixing the active layers with the ratios and concentrations suggested by the BO. Despite the much faster GPR loop, LineOne finalized all cells by evaporating the hole transport layer and top electrode and determined experimentally measured efficiencies in parallel to the GPR prediction. LineOne operates film processing and device processing in parallel. The BO is initially using predicted PCE values until experimental values are available.

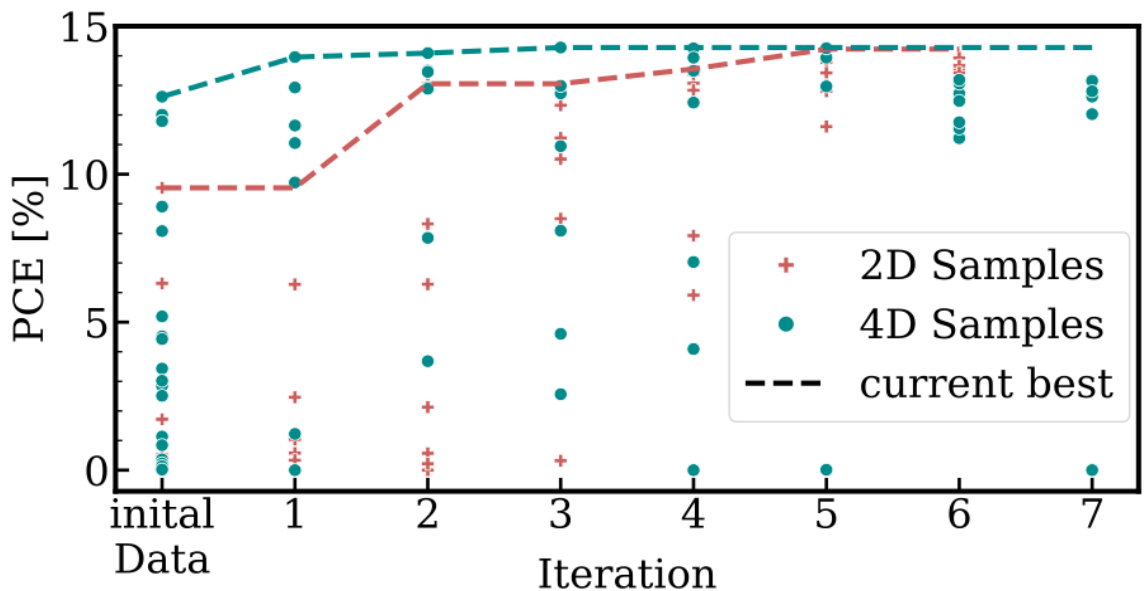

**Figure 2: PCE of the cells during the optimization runs**. Each point represents the experimentally measured PCE of the best cell of one sample fabricated during the optimization run for the 2D (red) and 4D (green) experiments. The dotted lines show the highest efficiency of all devices until the respective iteration. The initial data consisted of 7 (2D) and 21 (4D) samples. In each iteration, seven new Samples were fabricated. Therefore the best sample was found after 42 samples (5 Iterations) in the 2D optimization and 42 samples (3 Iterations) in the 4D optimization.

Figure 2 shows that the best-performing cells are identified after a few iterations in both experiments, yielding a PCE of 14.2 % and 14.3 % for the 2D and 4D experiments, respectively. Note that for the initial data set from LHS, seven substrates were used for the 2D optimization and 21 substrates for the 4D optimization. Therefore, only 42 samples were necessary to find the optimized parameter set in both experiments. To quantify the algorithm's performance, we run the BO 50 times on simulated functions obtained by a Gaussian process fit of the experimental data (Fig. S1 a). On average, it takes around four iterations to find the best cell in two dimensions, while it takes eight in four dimensions, corresponding to 35 and 77 samples, respectively. In comparison, a grid search with only ten steps per parameter would already require 91 substrates in two and 9.100 substrates in four dimensions.

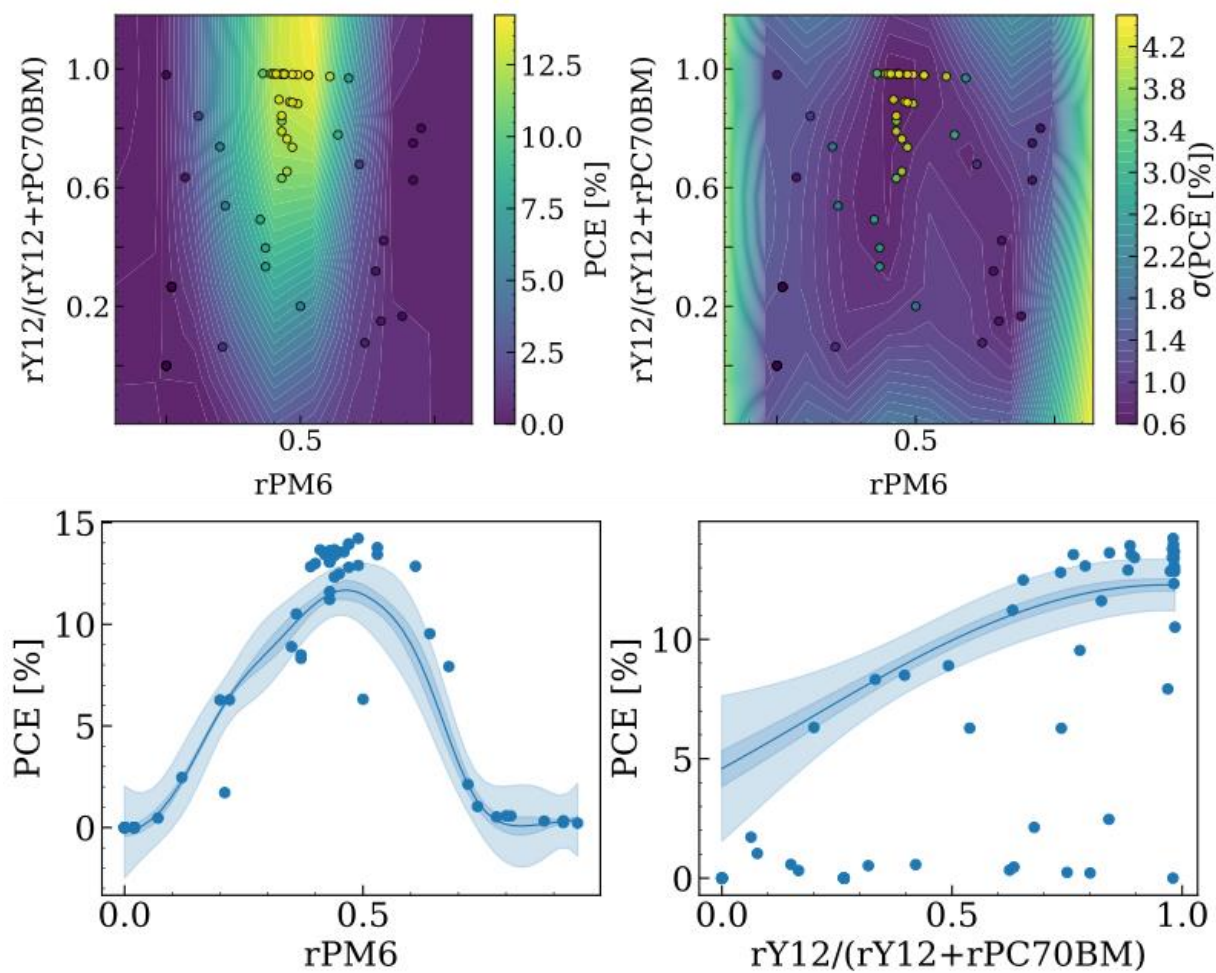

**Figure 3: Ratio Optimization of PM6:Y12:PC70BM.** (a) The approximate objective function and (b) the uncertainty of the optimization run using a Gaussian Progression Regression with a 5/2 Matern Kernel. The open parameters rPM6 (ratio of PM6 in the total system) and rY12/(rY12+rPC70BM) (ratio of Y12 normalized to the Acceptor share) are shown on the x and y-axis, respectively. The color of the points shows the measured efficiencies of the samples. (c) 1D partial dependence plots were obtained by running the model 500 times using bootstrapping. The solid lines show the mean of the models, the dark blue area is the standard deviation of the different models, and the bright blue area is the confidence interval of one model. The best cells are produced using an equal donor-acceptor share. Also, the efficiency of the cells increases when lowering the PC70BM share in the samples. While the uncertainty of the model is low in the regions where multiple cells are fabricated, it increases in the regions with few cells.

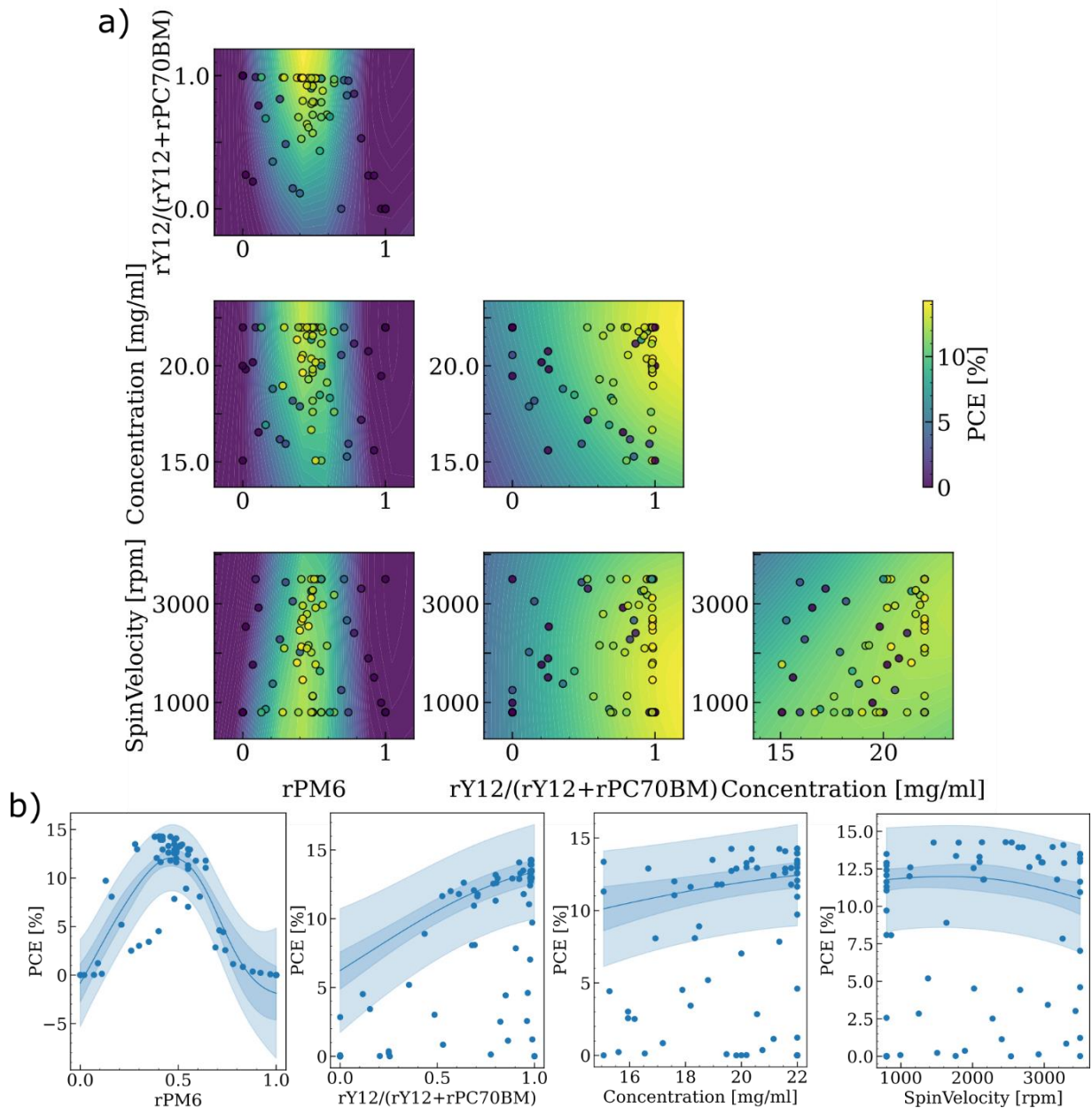

**Figure 4: Visualization of the influence of the experimentally controlled parameters on the efficiency during the ratio and process Optimization.** Shown are the four open parameters rPM6 (ratio of PM6 in the total system), rY12/(rY12+rPC70BM) (ratio of Y12 normalized to the Acceptor share), concentration [mg/mL], and the spin velocity [rpm]. (a) 2D partial dependence plots showing the approximated objective function of the model. (b) 1D partial dependence plots were obtained by running the model 500 times using bootstrapping. The solid lines show the mean of the models, the dark blue area is the standard deviation of the different models, and the bright blue area is the confidence interval of one model. The most dominant parameter is the ratio of PM6. Like in the 2D optimization, a donor-acceptor ratio close to 1:1 and a high amount of Y12 are favorable. Increasing the concentration leads to slightly better cells, while the spin Velocity does not significantly impact the performance.

Most importantly, we obtain a precise objective function (Figure 3 a), Figure 4 a)) over the whole design space. Visualizing the objective function helps us understand the coverage of the design space and the target response of our system, identifying regions of interest and ruling out non-performing parameter sets. Additionally, the BO returns the uncertainty of each point in the parameter space [35], [37], shown in Figure 2 b) and S6. Regions containing multiple samples show low uncertainty (around 0.5 % PCE), corresponding to the experimental reproducibility on LineOne (Fig. S2). In contrast, the optimizer has a higher uncertainty in regions with low numbers of samples. Since we obtained the estimated target response of all parameter sets, we can identify the importance of each parameter, allowing us to focus on the dominant parameter and avoid wasting resources on those with marginal influence in upcoming experiments. The partial dependence plots (PDP) in Figure 2 c) and Figure 3 b) visualize the effect of an individual feature of interest on the target response [51]. The model was run 500 times using bootstrapping. Based on the small variation in the dark blue area representing the standard deviation of the mean prediction across all runs, we can conclude that the model is robust and stable.

Additionally, the importance of each parameter in the model can be quantified by looking at the difference between the highest and lowest predicted efficiency values. Parameters significantly influencing the target response will show a large difference between these values. For such parameters, the PCE of the produced samples closely follows the mean prediction of the model. However, for less dominant parameters, the PCE values can differ significantly as the effects of another parameter dominate them. The dominant parameter for both experiments is the Donor share (rPM6). The best PCE is achieved around a Donor ratio of around 0.55 with efficiencies of over 14 %, while the efficiencies drastically reduce to zero for very high or very low Donor shares. Small PCE gains can be realized for the 4D experiment by increasing the concentration and tuning the spin velocity, which shows the best performance at around 2000 rpm.

Secondly, the share of Y12 of the total acceptor share significantly influences the cells' performance. Replacing PC70BM with Y12 continuously improves the performance of the cells. To our surprise, the BO on LineOne indicates that binary PM6:Y12 cells reach higher PCEs than the ternary ones. This is in contrast to reports in the literature highlighting the better performance of ternary OPV blends [52], [53]. However, we denote that LineOne operates under a constrained parameter space, which is atypical to manual optimization. Our data give insight that the absolute performance values of a device depend on the width of the parameter space allowed for optimization. On the one hand, research wants to keep the parameter space as small as possible to allow faster screening. On the other hand, as most parameters are not orthogonal to each other, a minimum dimension appears to be necessary. This underlines the importance of LineOne, which operates at outstanding reproducibility in well-defined dimensions. As an outlook to the future, we expect that systems like LineOne will be able to identify the most relevant and orthogonal processing parameters necessary to fully optimize novel material systems.

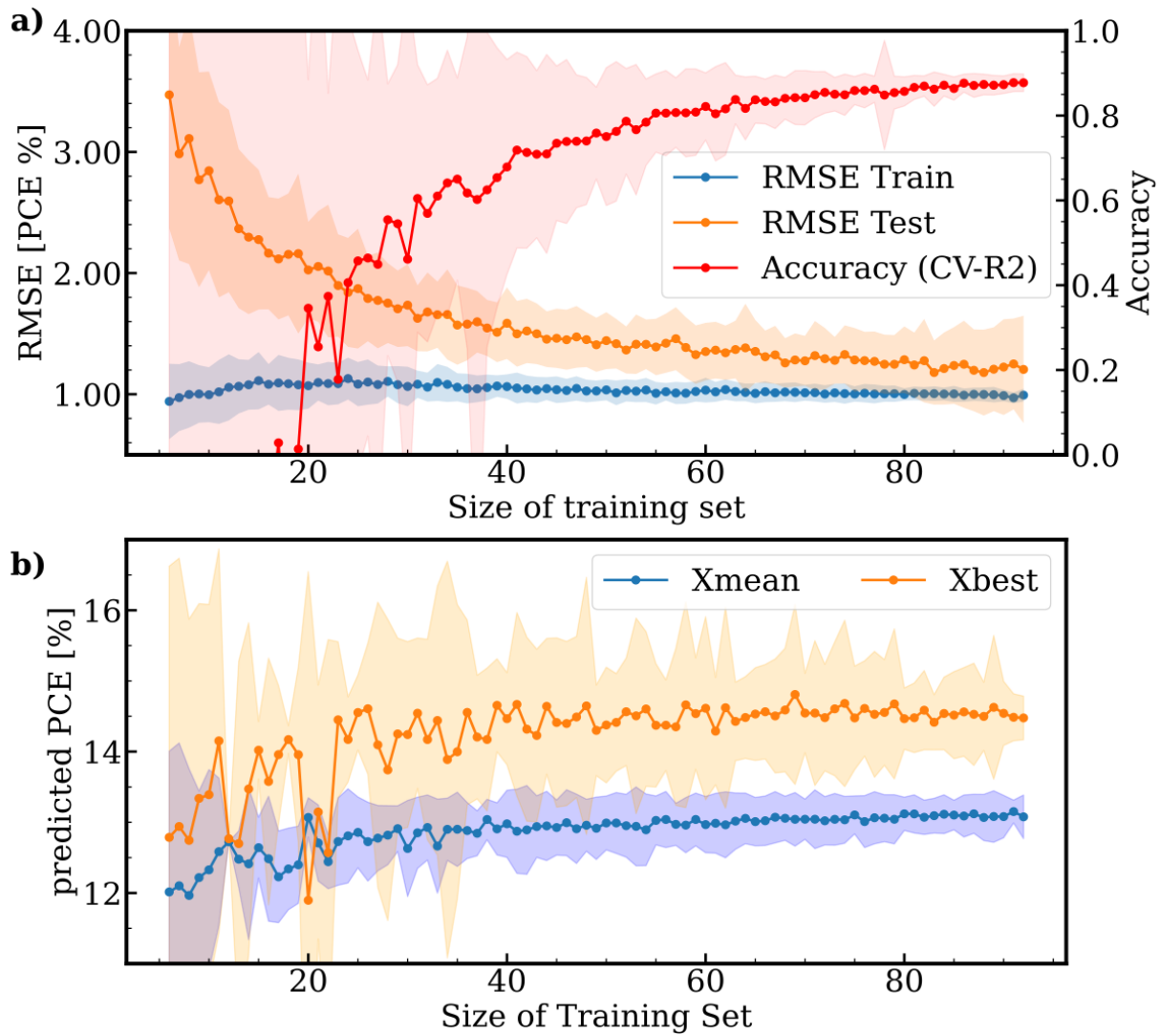

**Figure 5: GPR PCE Prediction Performance.** Investigation of the performance and number of samples necessary for good prediction of the PCE using the spectral features in a GPR were performed. a) The RMSE of the test and training set, the three-fold cross-validated R2 score (accuracy), and the respective standard deviation, depending on the number of samples in the training set, are shown. While the RMSE of the training set is constant over the whole range, the RMSE of the test set decreases, and the accuracy increases with more training data available. b) The predicted performance and 95 % confidence intervals are shown for the values of the features in the arithmetic mean (Xmean) and the parameter set with the highest predicted PCE when using 80 % of the available sample as training data (Xbest).

After the measured efficiency of a new batch of cells was available, we retrained the GPR with the new data to improve the accuracy of our prediction model. Figure 5 a) shows the root-mean-squared error (RMSE) of the training and test set, as well as the three-fold cross-validated R2 score (accuracy) in dependence on the number of samples in the training set. The data set includes 99 different samples after we excluded bad-performing cells. Devices were defined as bad-performing cells if the PCE was below 1 %. The RMSE of the test set and the accuracy improve quite fast at the beginning and increases only slowly from approximately 50 samples onward. At around 50 samples, we got reasonably low RMSE and satisfyingly high accuracy of the test set. To avoid over-fitting, we limited the boundaries of the kernel hyperparameter using our expert knowledge from previous simulation campaigns. Therefore, the RMSE of the training set does not change significantly. To further test the prediction power of our model, we took the mean of the optical features (Xmean) and the parameter set with the best-predicted performance using 80 % of the available data as training data (Xbest) and let each model predict the PCE of those parameter sets (Figure 5 b)). The predicted values for those two parameters do not change more than 5 % after only 30 samples.

Moreover, we are not approaching any boundaries of the data set, supporting the conclusion that we have identified the best parameter set in our system and do not need to increase the design space (see S4 and S5). While the RMSE of the test set is around 1.1 % PCE at 80 samples, which is slightly higher than the experimental error of 0.5 % PCE on LineOne (S2), the confidence interval for the prediction of Xmean is less than 0.5 % PCE. With enough data available, the model distinguishes the experimental error from the trend. Therefore, the confidence interval of a prediction becomes lower than the experimental error of the system. Overall the spectral model predicts the efficiency of the samples precisely, even with few data points available.

Overall, LineOne, combined with a BO and our early prediction GPR, is able to fully optimize novel material systems at an exceptionally low number of experiments and with error tolerances below the experimental uncertainty of LineOne. Moreover, the precisely controlled dimension of the parameter space opens the possibility of identifying the most decisive parameters controlling performance.

## Conclusion and Outlook

In this manuscript, we demonstrated for the first time a fully autonomous optimization cycle for optoelectronic devices as complex as organic solar cells. Our workflow on LineOne allows us to quickly optimize OPV devices while creating comprehensive data sets containing information about the entire design space and identifying important parameters, all with negligible human effort.

Understanding the influence and importance of various compositional and process parameters on the performance of OPV devices is crucial for the fast optimization of new materials and can significantly impact the rate at which the technology matures and becomes ready for commercialization. Understanding new materials faster and utilizing fewer quantities and experiments will lead to significantly better utilization of valuable resources.

Here we report for the first time a proof-of-concept study on a self-driving lab that uses artificial intelligence algorithms to effectively optimize fully functional OPV devices by controlling composition and process parameters. Bayesian optimization makes it possible to efficiently find optimal performance parameters in high dimensional spaces where human intuition begins to fail. Additionally, utilizing "learned" structure-property relationships allows us to perform a reliable GPR-based performance prediction even before a full device is finished with the top electrode. This makes it possible to perform rapid optimization cycles without losing any information by including the actual measured performance later to replace the initially predicted data. The absorption spectrum thus becomes an important proxy that contains the link to the structure-property relationship between the morphology and electronic performance characteristics since the underlying spectral model is based on physical properties. Unraveling the structure-property relationship will lead to a better understanding of the processes in the active layer and their influence on the device's performance.

The obtained data is stored in accordance with the FAIR principles [54] to ensure the reusability of the data by other research groups. Creating, storing, and sharing data sets covering the whole parameter space may enable research groups to use them in their ML models, extending our knowledge and accelerating the material discovery process even further. By combining different models, we envisage workflows that will allow to operate higher-dimensional parameter spaces, including structural or computational information of the semiconductors, which has the potential to create design rules for photovoltaic semiconductors without expensive experiments.

We believe that autonomous (automated and AI-guided) experimentation will significantly accelerate the discovery of new materials and material science in general and will play a major role in the future. This will shift the workflow of researchers from repetitive tasks and highly manual labor in the lab to designing experiments and evaluating the data.

LineOne is not limited to OPV but can optimize any thin film system. In the future, our research line and framework will as well optimize other target objectives, such as the operation stability of devices using in-Line degradation chambers or optical transmission of the devices. At the same time, the modular structure of our systems allows us to extend the closed loop and integrate other measurements or deposition methods.

## Device Fabrication

The Devices made in this publication were processed using the automated setups SpinBot and LineOne with an inverted ITO/ZnO/PM6:Y12:PC70BM/MoOx/Ag structure. Commercially available 25x25 mm glass substrates with patterned ITO were cleaned in an ultrasonic bath in DI-Water, Acetone, and IPA for 10 min each. Afterward, ZnO (N10) received from Avantama AG was ultrasonicated for 1 min, filtered through an 0.45 µm Polyamide (PA) filter, and automatically spin-coated and annealed for 30 min at 200°C using our SpinBot. Up to 96 samples were stacked in a carrier container and brought into LineOne. PM6 was obtained by Solarmer, Y12 by 1-Material, and PC70BM by Solenne BV. Stock Solutions of 20 mg/mL and 22 mg/mL in o-Xylene were prepared and brought into LineOne.

For the first experiment, only the composition was optimized (2 open parameters), while the composition, concentration, and spin velocity (4 open parameters) were optimized for the second one. The concentration and spin speed for the first experiment was set to 20 mg/mL and 1200 rpm, respectively. Table 1 shows the used parameter and parameter bounds.

Table 1: Description of the open Parameters

| Parameter name | Bounds | Description |
|---|---|---|
| rPM6 | [0,1) | PM6 (Donor) ratio in the total volume |
| rY12/(rY12+rPC70BM) | [0,1) | Y12 ratio in the total Acceptor (Y12+PC70BM) share |
| Conc. [mg/mL] | [15,22] | Concentration of the active layer solution |
| Spin velocity [rpm] | [800,3500] | Spin velocity for the application of the AL |

The active layer solution was created by pipetting and mixing the stock solutions using a four-channel pipetting robot with a modified gripper (Hamilton Starlet) and a well-plate on an automatic shaker. The solution was deposited on the substrate via spin coating. Afterward, all samples were annealed for 10 min at 120°C. The absorption spectra of the films were measured on six different spots using a SpectraMax M2 Platereader (Molecular devices) from 330 to 960 nm in 10 nm steps. 10 nm MoO3 for the HTL and 100 nm silver for the top electrode were thermally evaporated using a shadow mask, creating six devices on each substrate with an active area of 0.08 cm$^2$ each. Finally, the jV-Curves of the samples in the dark and under AM1.5G were measured using a SINUS-70 (Wavelabs) solar simulator.

# Acknowledgments

CJB and JH acknowledge funding through the European Commission via the projects EMERGE and CITYSOLAR as well as funds from the Bavarian Minsistry for Economy, Energy and Technology under the project ELF-PV. TH acknowledges funding through the DFG Project POPULAR. LL acknowledges funding through the DFG funded projects EXTRAORDINAIRE and BR4032/22. CJB acknowledges funding through the CRC 953 CARBON ALLOTROPES. The authors acknowledge Christian Berger for his work as software developer of the web interface and software framework of LineOne.

# References

[1] M. Riede, D. Spoltore, und K. Leo, „Organic Solar Cells—The Path to Commercial Success", *Adv. Energy Mater.*, Bd. 11, Nr. 1, S. 2002653, 2021, doi: 10.1002/aenm.202002653.

[2] M. Kaltenbrunner *u. a.*, „Ultrathin and lightweight organic solar cells with high flexibility", *Nat. Commun.*, Bd. 3, Nr. 1, Art. Nr. 1, Apr. 2012, doi: 10.1038/ncomms1772.

[3] C. J. M. Emmott *u. a.*, „Organic photovoltaic greenhouses: a unique application for semi-transparent PV?", *Energy Environ. Sci.*, Bd. 8, Nr. 4, S. 1317–1328, Apr. 2015, doi: 10.1039/C4EE03132F.

[4] R. Xia, C. J. Brabec, H.-L. Yip, und Y. Cao, „High-Throughput Optical Screening for Efficient Semitransparent Organic Solar Cells", *Joule*, Bd. 3, Nr. 9, S. 2241–2254, Sep. 2019, doi: 10.1016/j.joule.2019.06.016.

[5] F. Guo *u. a.*, „ITO-Free and Fully Solution-Processed Semitransparent Organic Solar Cells with High Fill Factors", *Adv. Energy Mater.*, Bd. 3, Nr. 8, S. 1062–1067, 2013, doi: 10.1002/aenm.201300100.

[6] P. Maisch *u. a.*, „Inkjet printed silver nanowire percolation networks as electrodes for highly efficient semitransparent organic solar cells", *Org. Electron.*, Bd. 38, S. 139–143, Nov. 2016, doi: 10.1016/j.orgel.2016.08.006.

[7] O. Almora *u. a.*, „Device Performance of Emerging Photovoltaic Materials (Version 3)", *Adv. Energy Mater.*, Bd. 13, Nr. 1, S. 2203313, 2023, doi: 10.1002/aenm.202203313.

[8] Z.-W. Zhao, Y. Geng, A. Troisi, und H. Ma, „Performance Prediction and Experimental Optimization Assisted by Machine Learning for Organic Photovoltaics", *Adv. Intell. Syst.*, Bd. 4, Nr. 6, S. 2100261, 2022, doi: 10.1002/aisy.202100261.

[9] J.-P. Correa-Baena *u. a.*, „Accelerating Materials Development via Automation, Machine Learning, and High-Performance Computing", *Joule*, Bd. 2, Nr. 8, S. 1410–1420, Aug. 2018, doi: 10.1016/j.joule.2018.05.009.

[10] B. Cao *u. a.*, „How To Optimize Materials and Devices via Design of Experiments and Machine Learning: Demonstration Using Organic Photovoltaics", *ACS Nano*, Bd. 12, Nr. 8, S. 7434–7444, Aug. 2018, doi: 10.1021/acsnano.8b04726.

[11] E. Maine und E. Garnsey, „Commercializing generic technology: The case of advanced materials ventures", *Res. Policy*, Bd. 35, Nr. 3, S. 375–393, Apr. 2006, doi: 10.1016/j.respol.2005.12.006.

[12]   J. Wagner, C. G. Berger, X. Du, T. Stubhan, J. A. Hauch, und C. J. Brabec, „The evolution of Materials Acceleration Platforms: toward the laboratory of the future with AMANDA", *J. Mater. Sci.*, Bd. 56, Nr. 29, S. 16422–16446, Okt. 2021, doi: 10.1007/s10853-021-06281-7.

[13]   R. Macarron *u. a.*, „Impact of high-throughput screening in biomedical research", *Nat. Rev. Drug Discov.*, Bd. 10, Nr. 3, Art. Nr. 3, März 2011, doi: 10.1038/nrd3368.

[14]   M. Liu *u. a.*, „High-Throughput Purification Platform in Support of Drug Discovery", *ACS Comb. Sci.*, Bd. 14, Nr. 1, S. 51–59, Jan. 2012, doi: 10.1021/co200138h.

[15]   W. H. Wang, X. Y. Liu, und Y. Sun, „High-Throughput Automated Injection of Individual Biological Cells", *IEEE Trans. Autom. Sci. Eng.*, Bd. 6, Nr. 2, S. 209–219, Apr. 2009, doi: 10.1109/TASE.2008.917136.

[16]   A. Buitrago Santanilla *u. a.*, „Nanomole-scale high-throughput chemistry for the synthesis of complex molecules", *Science*, Bd. 347, Nr. 6217, S. 49–53, Jan. 2015, doi: 10.1126/science.1259203.

[17]   J. J. de Pablo, B. Jones, C. L. Kovacs, V. Ozolins, und A. P. Ramirez, „The Materials Genome Initiative, the interplay of experiment, theory and computation", *Curr. Opin. Solid State Mater. Sci.*, Bd. 18, Nr. 2, S. 99–117, Apr. 2014, doi: 10.1016/j.cossms.2014.02.003.

[18]   A. Zakutayev *u. a.*, „An open experimental database for exploring inorganic materials", *Sci. Data*, Bd. 5, Nr. 1, Art. Nr. 1, Apr. 2018, doi: 10.1038/sdata.2018.53.

[19]   E. Soedarmadji, H. S. Stein, S. K. Suram, D. Guevarra, und J. M. Gregoire, „Tracking materials science data lineage to manage millions of materials experiments and analyses", *Npj Comput. Mater.*, Bd. 5, Nr. 1, Art. Nr. 1, Juli 2019, doi: 10.1038/s41524-019-0216-x.

[20]   J. O'Mara, B. Meredig, und K. Michel, „Materials Data Infrastructure: A Case Study of the Citrination Platform to Examine Data Import, Storage, and Access", *JOM*, Bd. 68, Nr. 8, S. 2031–2034, Aug. 2016, doi: 10.1007/s11837-016-1984-0.

[21]   L. Ward *u. a.*, „Matminer: An open source toolkit for materials data mining", *Comput. Mater. Sci.*, Bd. 152, S. 60–69, Sep. 2018, doi: 10.1016/j.commatsci.2018.05.018.

[22]   L. Ward, A. Agrawal, A. Choudhary, und C. Wolverton, „A general-purpose machine learning framework for predicting properties of inorganic materials", *Npj Comput. Mater.*, Bd. 2, Nr. 1, Art. Nr. 1, Aug. 2016, doi: 10.1038/npjcompumats.2016.28.

[23]   T. Ueno, T. D. Rhone, Z. Hou, T. Mizoguchi, und K. Tsuda, „COMBO: An efficient Bayesian optimization library for materials science", *Mater. Discov.*, Bd. 4, S. 18–21, Juni 2016, doi: 10.1016/j.md.2016.04.001.

[24]   L. M. Roch *u. a.*, „ChemOS: An orchestration software to democratize autonomous discovery", *PLOS ONE*, Bd. 15, Nr. 4, S. e0229862, Apr. 2020, doi: 10.1371/journal.pone.0229862.

[25]   R. Ramprasad, R. Batra, G. Pilania, A. Mannodi-Kanakkithodi, und C. Kim, „Machine learning in materials informatics: recent applications and prospects", *Npj Comput. Mater.*, Bd. 3, Nr. 1, Art. Nr. 1, Dez. 2017, doi: 10.1038/s41524-017-0056-5.

[26]   K. Roy, S. Kar, und D. Rudra Narayan, „Understanding the Basics of QSAR for Applications in Pharmaceutical Sciences and Risk Assessment - 1st Edition", 2015. https://www.elsevier.com/books/understanding-the-basics-of-qsar-for-applications-in-pharmaceutical-sciences-and-risk-assessment/roy/978-0-12-801505-6 (zugegriffen 22. November 2022).

[27]   S. Kar, N. Sizochenko, L. Ahmed, V. S. Batista, und J. Leszczynski, „Quantitative structure-property relationship model leading to virtual screening of fullerene derivatives: Exploring structural attributes critical for photoconversion efficiency of polymer solar cell acceptors", *Nano Energy*, Bd. 26, S. 677–691, Aug. 2016, doi: 10.1016/j.nanoen.2016.06.011.

[28]   M.-H. Lee, „Insights from Machine Learning Techniques for Predicting the Efficiency of Fullerene Derivatives-Based Ternary Organic Solar Cells at Ternary Blend Design", *Adv. Energy Mater.*, Bd. 9, Nr. 26, S. 1900891, 2019, doi: 10.1002/aenm.201900891.

[29]   Y. Wu, J. Guo, R. Sun, und J. Min, „Machine learning for accelerating the discovery of high-performance donor/acceptor pairs in non-fullerene organic solar cells", *Npj Comput. Mater.*, Bd. 6, Nr. 1, Art. Nr. 1, Aug. 2020, doi: 10.1038/s41524-020-00388-2.

[30]   Y. Cui *u. a.*, „Over 16% efficiency organic photovoltaic cells enabled by a chlorinated acceptor with increased open-circuit voltages", *Nat. Commun.*, Bd. 10, Nr. 1, Art. Nr. 1, Juni 2019, doi: 10.1038/s41467-019-10351-5.

[31]   X. Du *u. a.*, „Elucidating the Full Potential of OPV Materials Utilizing a High-Throughput Robot-Based Platform and Machine Learning", *Joule*, Bd. 5, Nr. 2, S. 495–506, Feb. 2021, doi: 10.1016/j.joule.2020.12.013.

[32]   Y. Shi, P. L. Prieto, T. Zepel, S. Grunert, und J. E. Hein, „Automated Experimentation Powers Data Science in Chemistry", *Acc. Chem. Res.*, Bd. 54, Nr. 3, S. 546–555, Feb. 2021, doi: 10.1021/acs.accounts.0c00736.

[33]     M. Seifrid u. a., „Autonomous Chemical Experiments: Challenges and Perspectives on Establishing a Self-Driving Lab", *Acc. Chem. Res.*, Bd. 55, Nr. 17, S. 2454–2466, Sep. 2022, doi: 10.1021/acs.accounts.2c00220.

[34]     M. Ahmadi, M. Ziatdinov, Y. Zhou, E. A. Lass, und S. V. Kalinin, „Machine learning for high-throughput experimental exploration of metal halide perovskites", *Joule*, Bd. 5, Nr. 11, S. 2797–2822, Nov. 2021, doi: 10.1016/j.joule.2021.10.001.

[35]     A. Gelman, J. B. Carlin, H. S. Stern, und D. B. Rubin, *Bayesian Data Analysis*. New York: Chapman and Hall/CRC, 1995. doi: 10.1201/9780429258411.

[36]     J. Kruschke, *Doing Bayesian Data Analysis: A Tutorial with R, JAGS, and Stan*. Academic Press, 2014.

[37]     O. Martin, *Bayesian Analysis with Python: Introduction to statistical modeling and probabilistic programming using PyMC3 and ArviZ, 2nd Edition*. Packt Publishing Ltd, 2018.

[38]     B. Lambert, *A Student's Guide to Bayesian Statistics*. SAGE, 2018.

[39]     Z. Ghahramani, „Probabilistic machine learning and artificial intelligence", *Nature*, Bd. 521, Nr. 7553, Art. Nr. 7553, Mai 2015, doi: 10.1038/nature14541.

[40]     F. Strieth-Kalthoff, F. Sandfort, M. Kühnemund, F. R. Schäfer, H. Kuchen, und F. Glorius, „Machine Learning for Chemical Reactivity: The Importance of Failed Experiments", *Angew. Chem. Int. Ed.*, Bd. 61, Nr. 29, S. e202204647, 2022, doi: 10.1002/anie.202204647.

[41]     W. Beker u. a., „Machine Learning May Sometimes Simply Capture Literature Popularity Trends: A Case Study of Heterocyclic Suzuki–Miyaura Coupling", *J. Am. Chem. Soc.*, Bd. 144, Nr. 11, S. 4819–4827, März 2022, doi: 10.1021/jacs.1c12005.

[42]     B. Burger u. a., „A mobile robotic chemist", *Nature*, Bd. 583, Nr. 7815, S. 237–241, Juli 2020, doi: 10.1038/s41586-020-2442-2.

[43]     B. P. MacLeod u. a., „A self-driving laboratory advances the Pareto front for material properties", *Nat. Commun.*, Bd. 13, Nr. 1, Art. Nr. 1, Feb. 2022, doi: 10.1038/s41467-022-28580-6.

[44]     C. W. Coley u. a., „A robotic platform for flow synthesis of organic compounds informed by AI planning", *Science*, Bd. 365, Nr. 6453, S. eaax1566, Aug. 2019, doi: 10.1126/science.aax1566.

[45]     A. Dave, J. Mitchell, S. Burke, H. Lin, J. Whitacre, und V. Viswanathan, „Autonomous optimization of non-aqueous Li-ion battery electrolytes via robotic experimentation and

machine learning coupling", *Nat. Commun.*, Bd. 13, Nr. 1, Art. Nr. 1, Sep. 2022, doi: 10.1038/s41467-022-32938-1.

[46]     P. M. Attia *u. a.*, „Closed-loop optimization of fast-charging protocols for batteries with machine learning", *Nature*, Bd. 578, Nr. 7795, Art. Nr. 7795, Feb. 2020, doi: 10.1038/s41586-020-1994-5.

[47]     S. Langner *u. a.*, „Beyond Ternary OPV: High-Throughput Experimentation and Self-Driving Laboratories Optimize Multicomponent Systems", *Adv. Mater.*, Bd. 32, Nr. 14, S. 1907801, 2020, doi: 10.1002/adma.201907801.

[48]     B. P. MacLeod *u. a.*, „Self-driving laboratory for accelerated discovery of thin-film materials", *Sci. Adv.*, Bd. 6, Nr. 20, S. eaaz8867, Mai 2020, doi: 10.1126/sciadv.aaz8867.

[49]     T. Head *u. a.*, „scikit-optimize/scikit-optimize: v0.5.2". Zenodo, 25. März 2018. doi: 10.5281/zenodo.1207017.

[50]     C. Chevalier und D. Ginsbourger, „Fast Computation of the Multi-Points Expected Improvement with Applications in Batch Selection", in *Learning and Intelligent Optimization*, G. Nicosia und P. Pardalos, Hrsg., in Lecture Notes in Computer Science. Berlin, Heidelberg: Springer, 2013, S. 59–69. doi: 10.1007/978-3-642-44973-4_7.

[51]     J. H. Friedman, „Greedy function approximation: A gradient boosting machine.", *Ann. Stat.*, Bd. 29, Nr. 5, S. 1189–1232, Okt. 2001, doi: 10.1214/aos/1013203451.

[52]     N. Y. Doumon, L. Yang, und F. Rosei, „Ternary organic solar cells: A review of the role of the third element", *Nano Energy*, Bd. 94, S. 106915, Apr. 2022, doi: 10.1016/j.nanoen.2021.106915.

[53]     P. Cheng *u. a.*, „Ternary System with Controlled Structure: A New Strategy toward Efficient Organic Photovoltaics", *Adv. Mater.*, Bd. 30, Nr. 8, S. 1705243, 2018, doi: 10.1002/adma.201705243.

[54]     M. D. Wilkinson *u. a.*, „The FAIR Guiding Principles for scientific data management and stewardship", *Sci. Data*, Bd. 3, Nr. 1, Art. Nr. 1, März 2016, doi: 10.1038/sdata.2016.18.

# Autonomous Optimization of an Organic Solar Cell in a 4-dimensional Parameter Space

# Supporting Information

## A  Full name of materials

**PM6:** Poly[(2,6-(4,8-bis(5-(2-ethylhexyl-3-fluoro)thiophen-2-yl)-benzo[1,2-b:4,5-b'] dithio-phene))-alt-(5,5-(1',3'-di-2-thienyl-5',7'-bis(2-ethylhexyl)benzo[1',2'-c:4',5'-c']dithiophene-4,8-dione)]

**Y12:**  2,2'-((2Z,2'Z)-((12,13-bis(2-butyloctyl)-3,9-diundecyl-12,13-dihydro-[1,2,5]thiadiazolo[3,4-e]thieno[2'',3''':4',5']thieno[2',3':4,5]pyrrolo[3,2-g]thieno[2',3':4,5]thieno[3,2-b]indole-2,10-diyl)bis(methanylylidene))bis(5,6-difluoro-3-oxo-2,3-dihydro-1H-indene-2,1-diylidene))dimalononitrile

**PC70BM:** [6,6]-Phenyl-C71-butyric acid methyl ester

# B    Model Evaluation Bayesian Optimizer

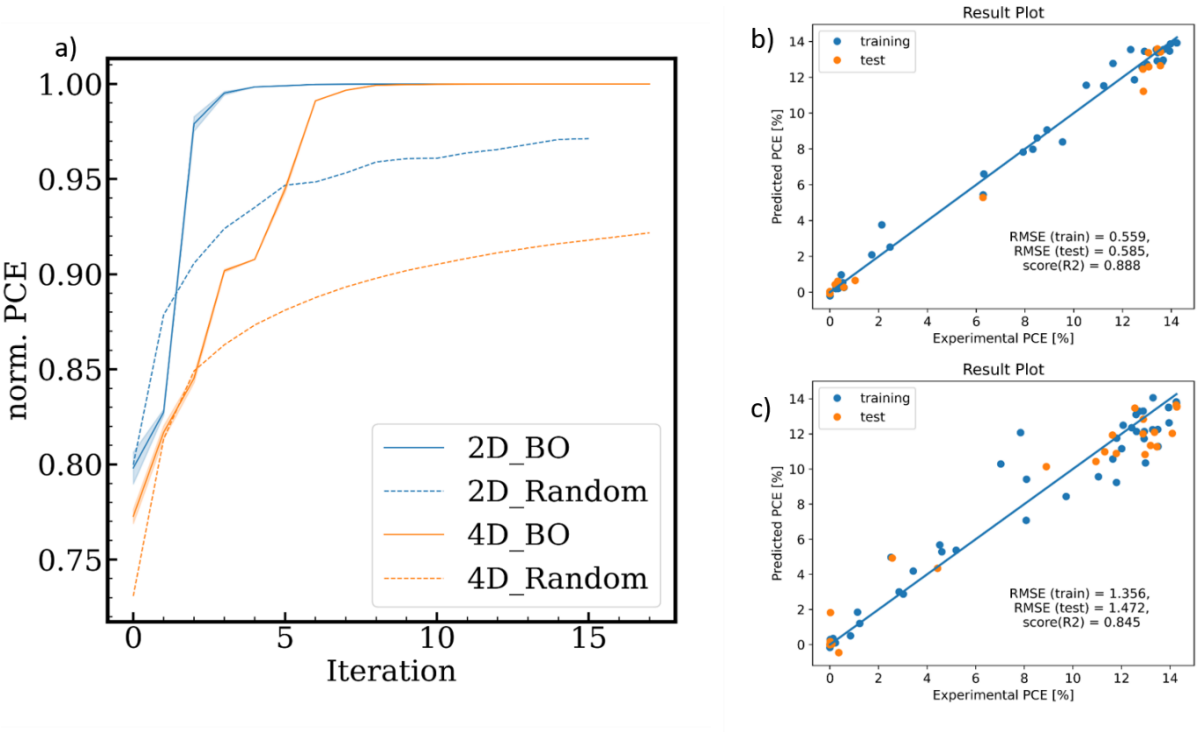

**Figure S1: Model Evaluation.** (a) The Bayesian Optimizer was run 50 times on the obtained objective function from the two optimizations. The solid lines (blue: 2D, orange: 4D) show the mean of the 50 runs of the cumulatively highest normalized efficiency of all devices until the respective iteration and the interquartile range is the shaded area. As a baseline, the dotted lines show the mean of 50 random sampling runs. The BO takes around five Iterations to find the optimum. Note that we obtained by LHS sampling the initial data sets of 7 and 21 samples for the 2D and 4D, respectively. The experimental vs. the predicted PCE of the model of the (b) 2D and (c) 4D optimization using a 70/30 training test split 50 times.

## C  LineOne Reproducibility of the Referenz Cells

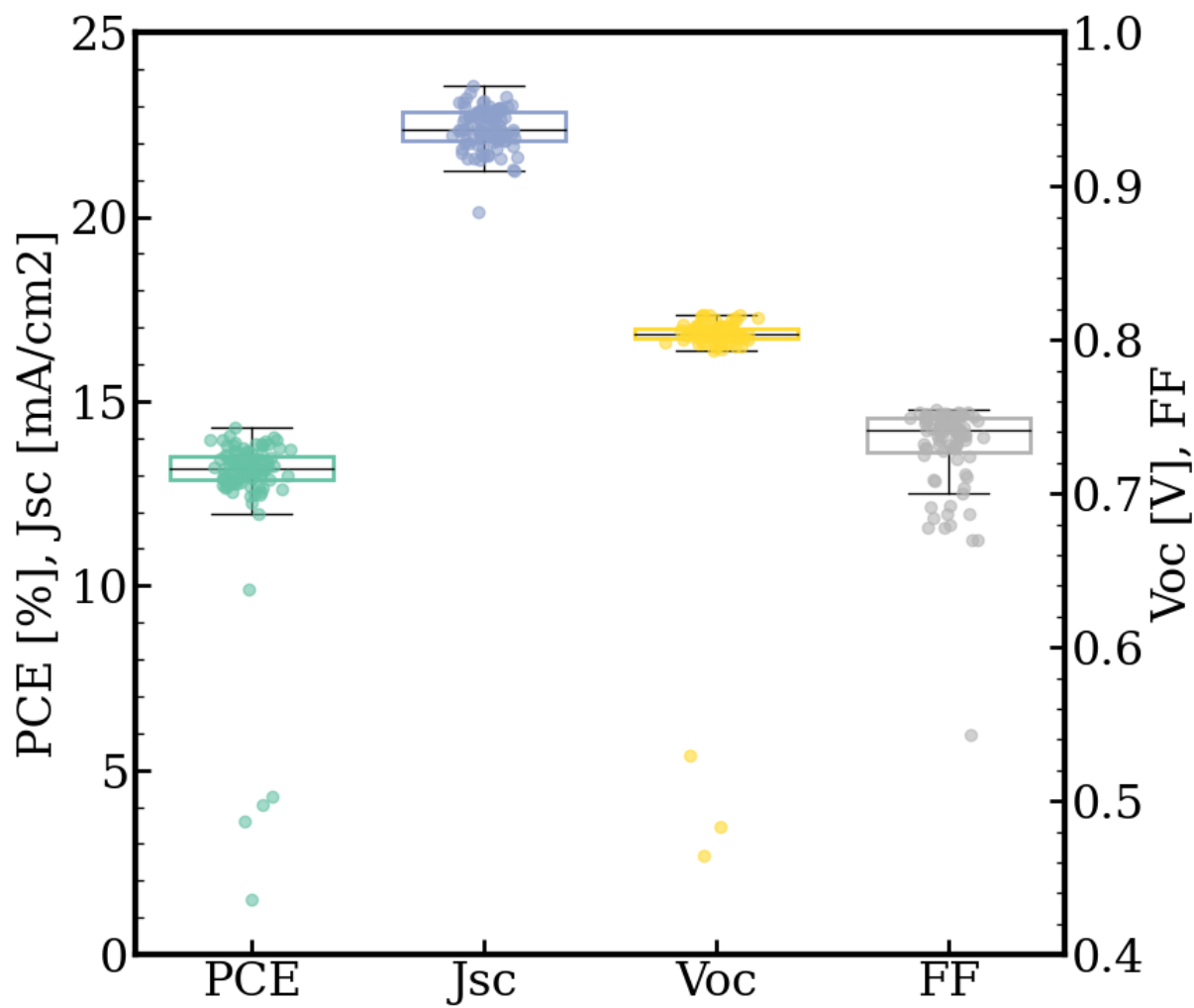

Figure S2: LineOne Reproducibility. Results of jV-measurements of the reference samples. In total 18 cells were fabricated with the same parameters (PM6:Y12:PC70BM (1:1.2:0.2), 20 mg/mL, 1200 rpm) during the experiments. Each point represents one cell (6 cells per sample).

# D     Model Performance of the GPR to predict the cell efficiency

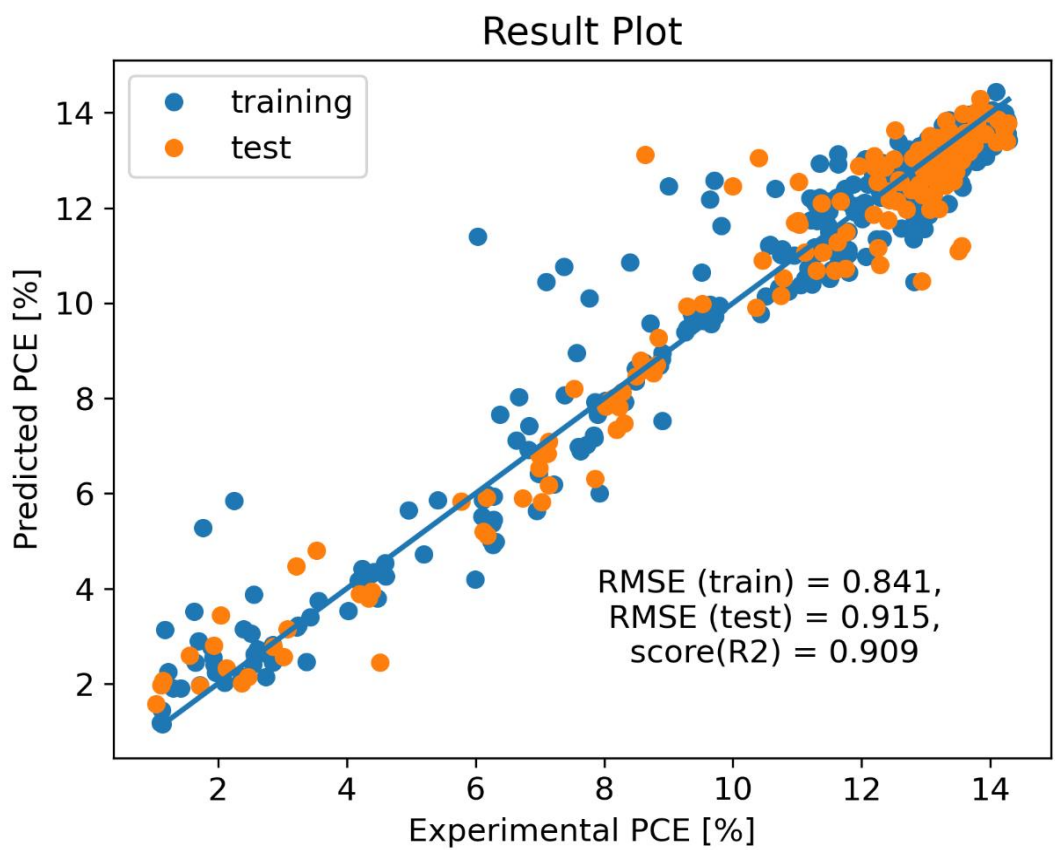

Figure S3: Model Performance of the PCE prediction using the UV-Vis data. The experimental vs. the predicted efficiency of the GPR model using a 70/30 test-train split 50 times. Each point represents one cell of the samples fabricated during the experiments. Cells with PCE > 1 % were excluded.

# E   Optimal spectral Features

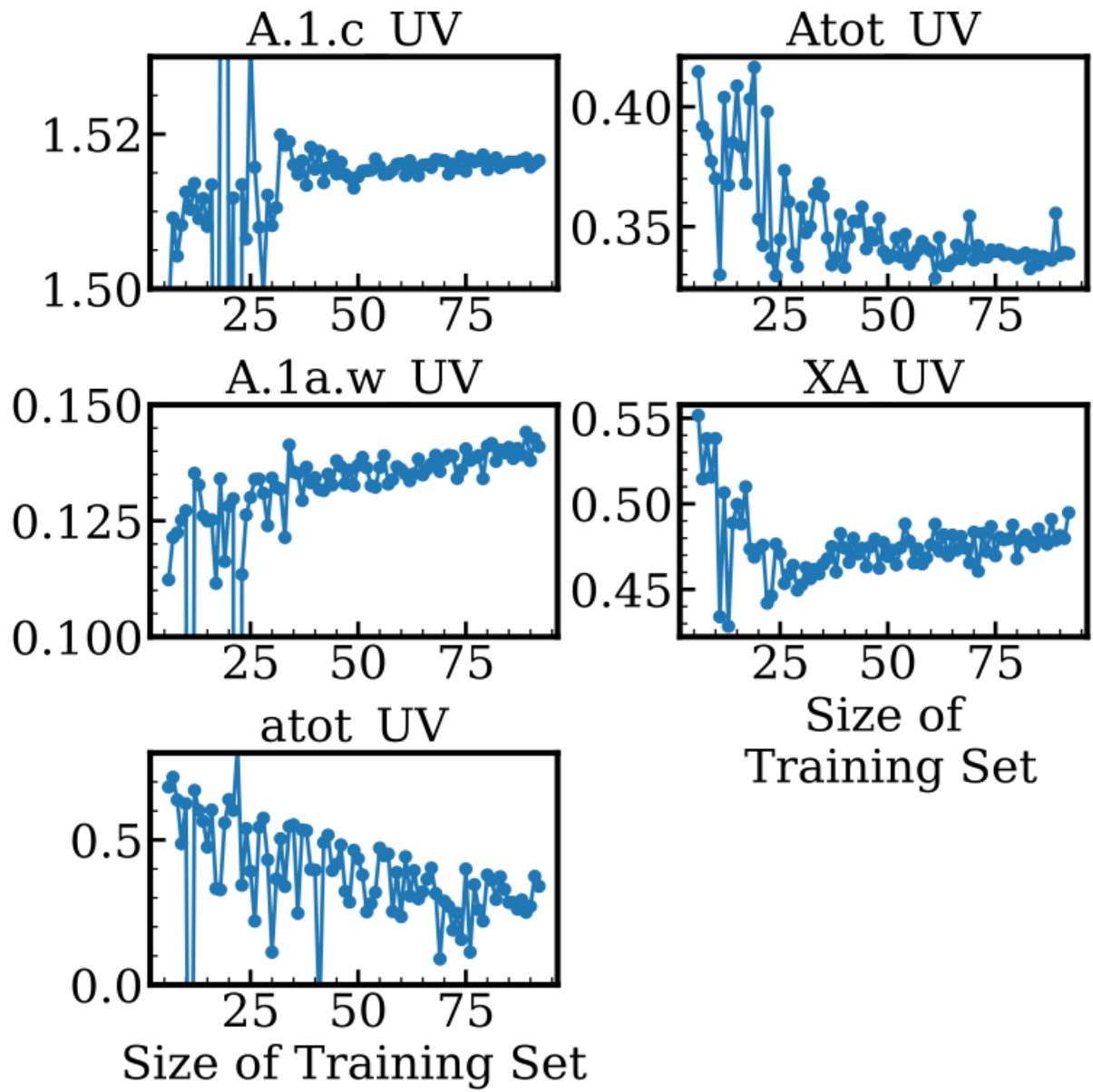

**Figure S4: Predicted Optimum of each optical feature depending on the size of the training set.**

# F  Distribution of the optical features in the dataset

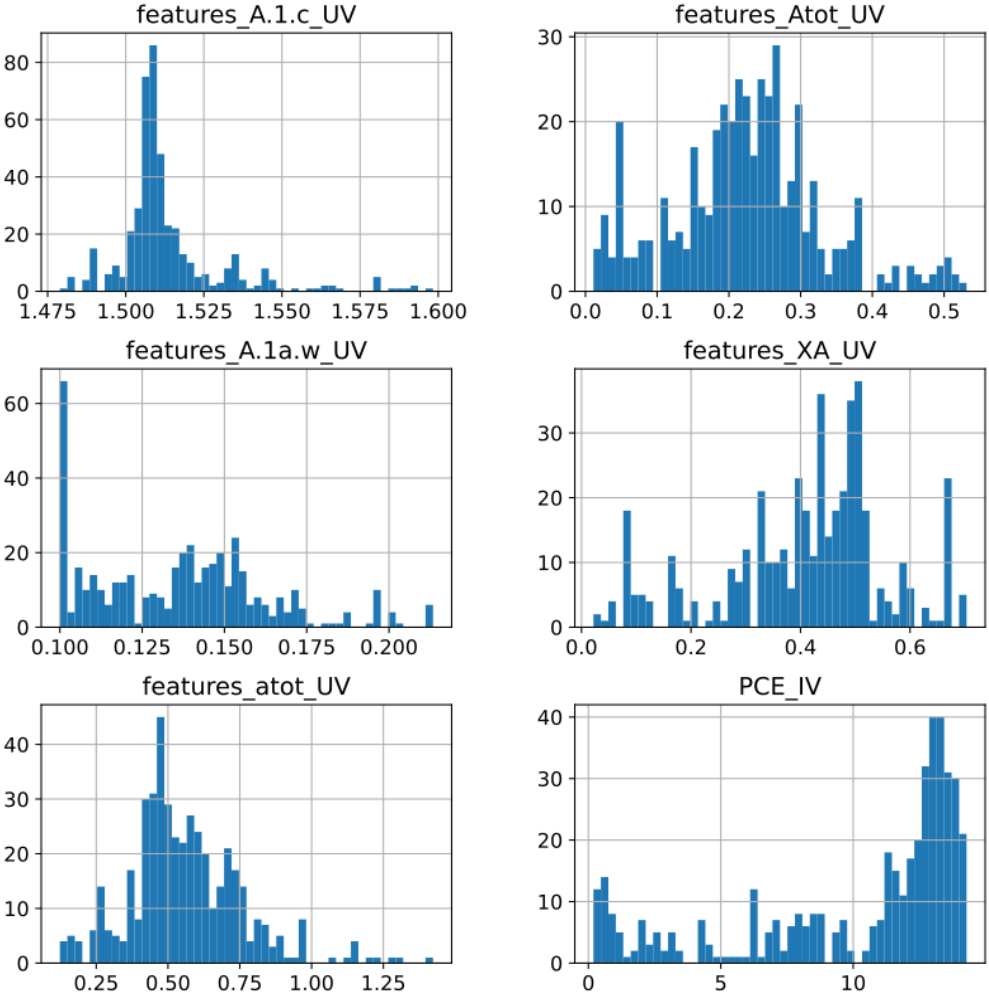

**Figure S5: Distribution of the spectral data**

# G     Uncertantiy of the Bayesian Optimizer in the 4D experiment

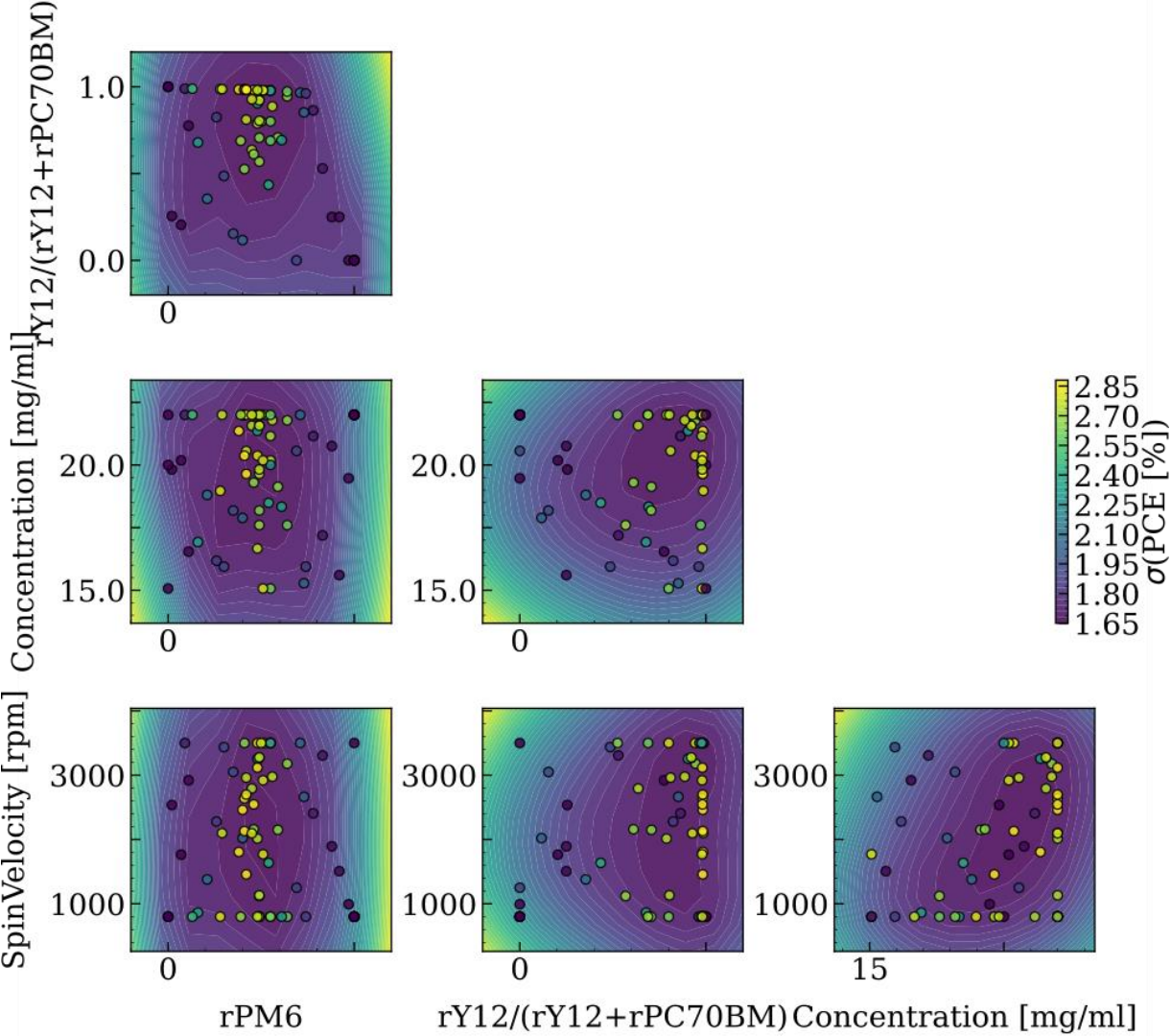

**Figure S6: Uncertainty of the Bayesian Optimizer during the 4D Experiment**